\documentclass{aa}  

\usepackage{graphicx}
\usepackage{txfonts}
\usepackage{float}

\begin{document}

   \title{Three-dimensional structure of the magnetic field \\ in the disk of the Milky Way}
   \titlerunning{Three-Dimensional Structure of the GMF}

   \subtitle{}

   \author{A. Ordog\inst{1}
          \and
           J.C. Brown\inst{1}
           \and
           R. Kothes\inst{2}
           \and
           T.L. Landecker\inst{2}
          }

   \institute{Department of Physics and Astronomy, University of Calgary,
    Calgary, Alberta, T2N 1N4, Canada\\
              \email{aordog@ucalgary.ca}
         \and
             Dominion Radio Astrophysical Observatory, Herzberg Programs in Astronomy and Astrophysics, National Research Council Canada, PO Box 248, Penticton, BC V2A 6J9, Canada\\
             }

\date{}

\abstract
  {} 
   {We present Rotation Measures (\textbf{RM}) of the diffuse Galactic synchrotron emission from the Canadian Galactic Plane Survey (\textbf{CGPS}) and compare them to RMs of extragalactic sources in order to study the large-scale reversal in the Galactic magnetic field (\textbf{GMF}).}
   {Using Stokes $Q$, $U$ and $I$ measurements of the Galactic disk collected with the Synthesis Telescope at the Dominion Radio Astrophysical Observatory, we calculate RMs over an extended region of the sky, focusing on the low longitude range of the CGPS ($\ell=52^{\circ}$ to $\ell=72^{\circ}$).}
   {We note the similarity in the structures traced by the compact sources and the extended emission and highlight the presence of a gradient in the RM map across an approximately diagonal line, which we identify with the well-known field reversal of the Sagittarius-Carina arm. We suggest that the orientation of this reversal is a geometric effect resulting from our location within a GMF structure arising from current sheets that are not perpendicular to the Galactic plane, as is required for a strictly radial field reversal, but that have at least some component parallel to the disk. Examples of models that fit this description are the three-dimensional dynamo-based model of Gressel et al. \citeyearpar{gresseletal} and a Galactic scale Parker spiral \citep{akasofuhakamada}, although the latter may be problematic in terms of Galactic dynamics.}
   {We emphasize the importance of constructing three-dimensional models of the GMF to account for structures like the diagonal RM gradient observed in this dataset.}


\maketitle

\section{Introduction}

The Galactic magnetic field (\textbf{GMF}) is recognized as an essential constituent of the interstellar medium. The field lines in the Galactic disk are approximately aligned with the material spiral arms, and are typically modelled as logarithmic spirals. Estimates of the pitch angle vary between 0$^{\circ}$ \citep{vallee08} and -30$^{\circ}$ (which includes the halo field; \citealt{fauvet}), with the most commonly cited value being around -11.5$^{\circ}$ \citep{haverkornreview}. Furthermore, the pitch angle likely varies with radius \citep{vaneck}. 

The general consensus has the predominant direction of the large-scale field clockwise, as viewed from the North Galactic pole, with one known reversed region directed counterclockwise \citep{vaneck,simardnormandinkronberg,thomsonnelson}. By Amp\`ere's law, such a magnetic ``shear'', or more commonly ``reversal'', requires the presence of a current sheet at the interface between the two magnetic regions. In a galactic disk, if the direction of the field is purely a function of radius, a magnetic field reversal would imply a current sheet perpendicular to the disk, for which two-dimensional modelling is sufficient. 

To date, there is no satisfactory explanation of the source mechanism for a large-scale reversal \citep{beck16}, nor a resolution to the question of how many reversals exist in the Galaxy. Some models suggest a single reversal \citep{vaneck}, while another suggests as many as one at each arm-interarm boundary \citep{hanetal}. Alternatively, the observed reversal may not be a large-scale feature, but rather a close-up view of a more local effect \citep{shukurov}.  Further complicating the problem is the fact that large-scale reversals are not observed in external galaxies \citep{beck16}.

We examine a previously un-noted characteristic of the large-scale reversal in the GMF in Rotation Measures (\textbf{RM}) of the extended polarised emission data from the Canadian Galactic Plane Survey (\textbf{CGPS}) and discuss its implications for understanding the structure of the field reversal. In doing so, we also highlight the usefulness of this particular dataset in the context of extracting information about GMF structures.

\section{The data}

When a linearly polarised electromagnetic wave passes through a magnetized electron gas such as the interstellar medium, its plane of polarisation rotates, an effect known as ``Faraday Rotation'':

\begin{equation}
\Delta \phi = \phi-\phi_{\circ} = \lambda^2(0.812 \int n_e \boldsymbol{B} \cdot d\boldsymbol{l}) = \lambda^2RM \hspace{0.1cm} [{\rm{rad}}].
\end{equation}

\noindent Here $\phi_{\circ}$ and $\phi$ are the polarisation angles at the source and observer respectively, $\lambda$ is the wavelength, $n_e$ is the electron density, $d\boldsymbol{l}$ is the path length increment along the line of sight (\textbf{LOS}) directed from the source to the observer, and $\boldsymbol{B}$ is the magnetic field in the region. The integral defines the Rotation Measure (RM) and can be determined for a given polarised source as the slope of $\phi$ versus $\lambda^2$. Positive (negative) RMs indicate an average magnetic field pointing toward (away from) the observer. RMs for many lines of sight can therefore be used to probe the GMF.

We use data from the CGPS \citep{tayloretal}, collected using the Synthesis Telescope at the Dominion Radio Astrophysical Observatory (\textbf{DRAO}), which has four, 7.5 MHz bands within a 35 MHz window, centred at 1420 MHz \citep{landecker10}. Simultaneous observations of Stokes $I$, $Q$ and $U$ allow for unambiguous determination of  RMs. The CGPS has the highest density of compact extragalactic (\textbf{EG}) source RMs in the Galactic disk to date \citep{brown03b}, with more than 1 source per square degree. However, little exploration of the CGPS extended emission (\textbf{XE}; diffuse synchrotron emission) RMs has been carried out. These data are presented in Figs. \ref{fig1} and \ref{fig2}.

The XE RM dataset has two possible limitations. First, while EG sources dominate the emission along their lines of sight, the XE originates from a range of depths, leading to significant depth depolarisation, whereby polarisation of light emitted at different points along the LOS (experiencing different amounts of Faraday rotation) averages out. Beyond a distance known as the Polarisation Horizon (\textbf{PH}; \citealt{uyaniker,kotheslandecker}; Kothes et al., in prep.), polarised emission from the XE is not detectable. Even within the PH, XE can be Faraday-thick if both emission and rotation occur within the same volume, resulting in different amounts of Faraday Rotation for emission produced at different distances \citep{burn,gardnerwhiteoak,sokoloff,carretti}. This in turn can lead to nonlinearities in polarisation angle as a function of $\lambda^2$, potentially making simple RM calculations unreliable. Rotation Measure Synthesis \citep{brentjensdebruyn} has been used to gain information on the Faraday depth structure of XE or other sources having multiple Faraday components, but the CGPS observations do not have the required dense sampling of $\lambda^2$ space for this technique to be employed.

Second, the synthesis array that obtained these data is inherently limited to measuring structures on angular scales smaller than about 30 arcminutes \citep{landecker10}. Our data are less sensitive to the larger polarisation structures that are part of the diffuse Galactic emission, and complementary single-antenna data are required to incorporate this information into the images. Single-antenna data from the John A. Galt telescope are available for total and polarised intensity (\textbf{PI}; see Fig. \ref{fig1}, 2nd and 3rd panels), but not in the four separate bands required for RM calculations \citep{landecker10}.

We counter these concerns as follows. First, if we observe some degree of spatial structure in the RM maps obtained using the assumption of a linear relationship between the polarisation angle and the wavelength squared, then it is reasonable to assume that there is meaningful information in those RMs. They may be interpreted as characteristic RMs of the lines of sight for the given wavelength range even if they lack detail about the full Faraday structure \citep{farnsworth}. Furthermore, the PH that limits the LOS distance makes the XE RMs ideal for probing the local field structure.

Although the XE RM data miss the largest angular structures, the interferometer is sensitive to high spatial frequencies, and hence detects sharp gradients across angular scales, such as that produced by a large-scale field reversal. Therefore, although we cannot be certain of where the zero levels of emission are in the Stokes $Q$ and $U$ maps used for calculating RMs, we can see where sharp changes in RM (and therefore in the magnetic field) occur, and we argue that this is valuable information.

\section{Analysis}
\begin{figure*}
\centering
   \includegraphics[width=15.9cm]{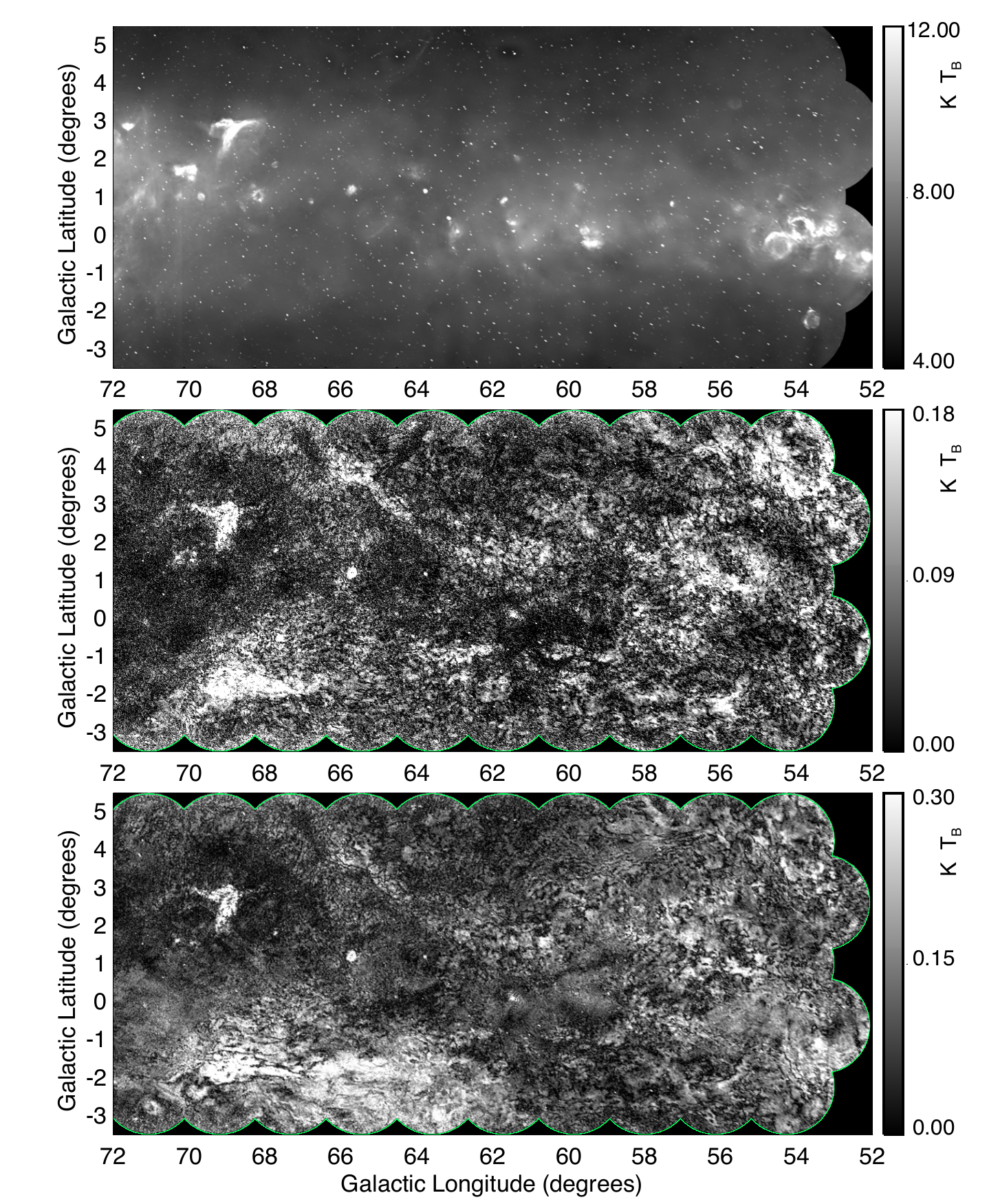}
     \caption{Low longitude CGPS intensity maps. \textit{Top panel}: Stokes $I$ from the DRAO interferometer data combined with the John A. Galt 26-m telescope data. \textit{Second panel}: Polarised intensity from the interferometer data only. \textit{Third panel}: Polarised intensity from the interferometer and single-antenna data combined. }
     \label{fig1}
\end{figure*}

For this analysis, we focus on the lowest 20$^{\circ}$ longitude range of the CGPS, ($52^{\circ}<\ell<72^{\circ}$; Figs. \ref{fig1} and \ref{fig2}), where we are looking toward the inner Galaxy, into the Sagittarius-Carina arm. Here the magnetic field component orthogonal to the LOS is sufficient to produce observable polarised synchrotron emission (Fig. \ref{fig1}, 2nd and 3rd panels) and the parallel component is sufficient to produce relatively large RMs in both the EG point sources (Fig. \ref{fig2}, 1st panel) and the XE (Fig. \ref{fig2}, 2nd panel). The RMs for both the XE and EG sources are calculated using Eq. (1), and any data points with a signal to noise ratio less than 5 for the XE RMs are discarded. The top panel of Fig. \ref{fig1} shows the total intensity map for comparison.

We note that despite the potential drawbacks outlined above, there is a significant degree of spatial structure present in the XE RM map, indicating that this is a useful dataset to study. Furthermore, we observe a striking degree of similarity between the PI map without single-antenna data (Fig. \ref{fig1}, 2nd panel) and the PI map including single-antenna data (Fig. \ref{fig1}, 3rd panel), indicating that much of the structure corresponds to the higher spatial frequencies that are observable with the DRAO Synthesis Telescope. This lends support to the reliability of interferometer-only data in this region.

\subsection{Observed RM distribution}
\begin{figure*}[t]
\centering
   \includegraphics[width=16.9cm]{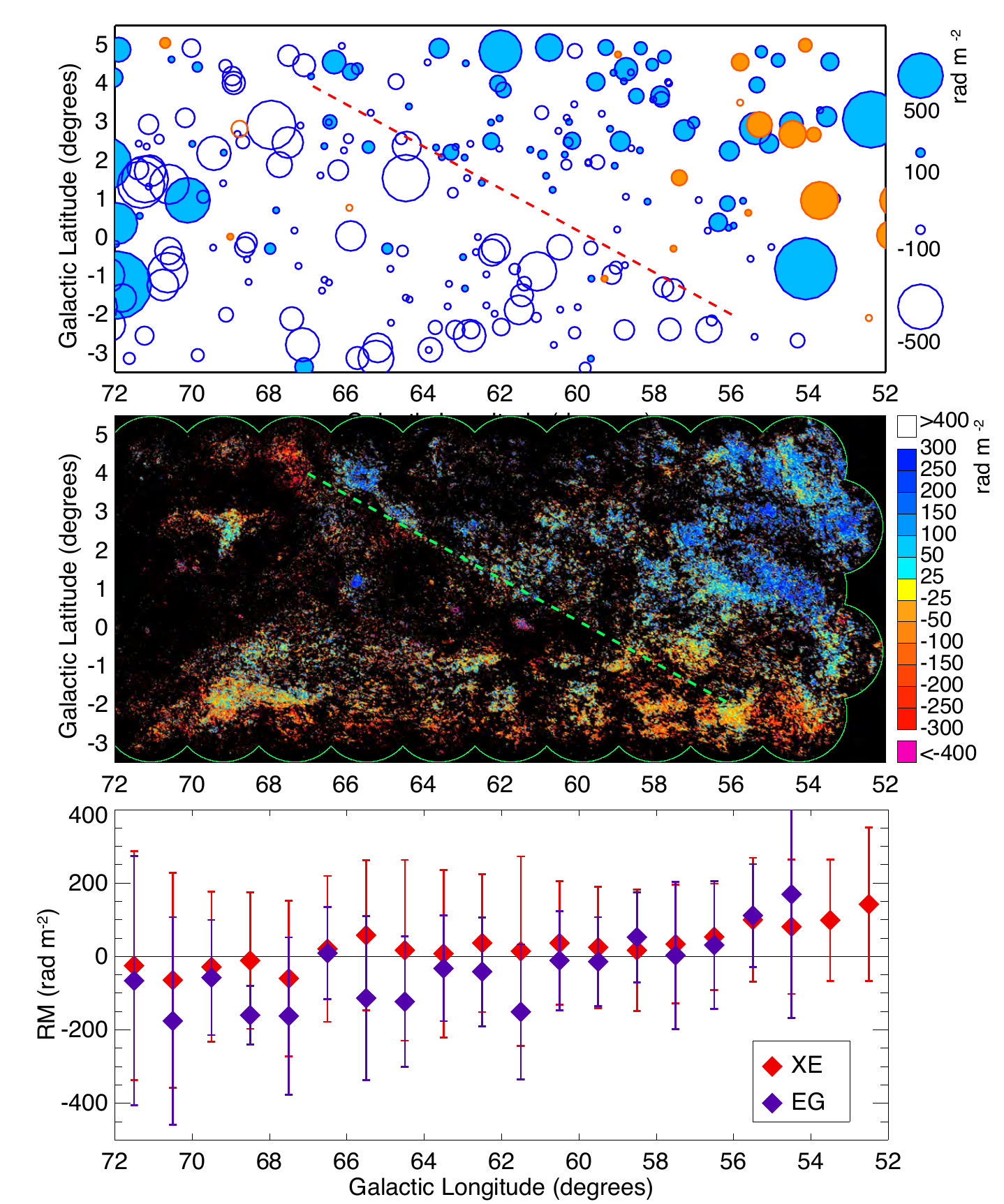}
     \caption{Low longitude CGPS Rotation Measures. \textit{Top panel}: RMs of extragalactic sources (blue) and pulsars (orange). \textit{Second panel}: RMs of extended emission using only interferometer data. The diagonal lines indicate the approximate location of the boundary across which the RM gradient occurs. \textit{Third panel}: RMs of extended emission sources (red) and extragalactic sources (purple) as a function of Galactic longitude, averaged into 1$^{\circ}$ longitude bins. The error bars were determined as the standard deviations within the bins.}
     \label{fig2}
\end{figure*}

The most remarkable feature of these maps is the gradient in RM across a diagonal boundary, observable in both the EG RMs and the XE RMs. The boundary, identified by the dashed lines in Fig. \ref{fig2}, extends from around $\ell =67^{\circ}, b=4^{\circ}$, to $\ell = 56^{\circ}, b=-2^{\circ}$, above which the RMs are predominantly positive and below which they are predominantly negative. This indicates that above (below) the boundary the LOS magnetic field is directed toward (away from) us, which we interpret as being counterclockwise (clockwise), viewed from the North Galactic pole. Comparing the EG RMs (1st panel) to the XE RMs (2nd panel), we observe that both RM tracers appear to follow a similar trend, although slightly higher magnitude RMs appear in the  EG sources than in the XE sources. This strong resemblance between RM maps derived from EG and XE data, over an extended area, has not been remarked on previously. Since the RMs agree so well in sign (and reasonably well in magnitude) we conclude that they are tracing the same magnetic field configuration, namely a large-scale field reversal.

\looseness=-1 The resemblance between the XE and EG RMs is surprising when we consider that the two sources likely probe very different spatial volumes. The assumption is that XE RMs probe the LOS only as far as the PH, which is closer than 2 kpc for \break $50^{\circ}<\ell<120^{\circ}$ (\citealt{uyaniker,kotheslandecker}; Kothes et al., in prep.). The EG RMs, on the other hand, probe the LOS out to the edge of the Galaxy. If this assumption is correct, one might expect the magnitudes of the EG RMs to be significantly larger than those of the XE RMs, depending on the field configuration. 

To comment on this, we examine the differences between the RMs of these two datasets by binning the data into 1$^{\circ}$ longitude bins\footnote{Bin sizes of 1$^{\circ}$ were chosen to ensure that a statistically significant number of EG sources fall within each bin.} and observing the variations between the two datasets as a function of longitude. The bottom panel of Fig. \ref{fig2} shows the binned XE and EG RMs, in which we can see the similarity in the general trends between the datasets. The EG RMs are indeed predominantly larger than their XE counterparts, but the largest difference is only around 150 rad m$^{-2}$. In an average electron density of 0.1 cm$^{-3}$ and a large-scale GMF that decreases as $r^{-1}$ from a local strength of 2 $\mu$G, this would occur over a distance of less than 1 kpc, much smaller than the difference in depths assumed to be probed by the EG and XE RMs.

\subsection{Interpretation}
\vspace{-0.05cm}
The small differences that do exist between the magnitudes of the EG and XE RM values are likely due to depth depolarisation, which is more significant for the XE than the EG sources. One possible explanation for the smaller-than-expected difference is that the magnetic field strength and electron density actually decay quickly enough with galactocentric distance beyond a few kpc so that the EG sources undergo most of their observed rotation nearby. A second possibility is that reversed regions of the magnetic field may also exist in the outer Galaxy and contribute to reducing the magnitudes of EG RMs so that they are comparable to the XE RMs that probe the nearby field. This could be verified by examining RMs of pulsars located between the XE PH and the edge of the Galaxy, but is beyond the scope of this paper and would ideally require a much higher density of pulsars with known distances than is presently available. The known pulsars in this region are shown in the top panel of Fig. \ref{fig2}, for comparison with the EG RMs. Their RMs are consistent with the pattern seen in the EG and XE datasets, but their paucity and the uncertainties in their distances do not allow definite conclusions about the number or location of field reversals.  A third possibility is that the PH near the reversal is considerably further than the assumed 2 kpc, since depth depolarisation reduces substantially in regions where the magnetic field drops to zero.

Four decades of RM observations have led to the interpretation of a magnetic field reversal between the local and Sagittarius-Carina arms. These observations would suggest a current sheet between the two spiral arms that is perpendicular to the disk. However, our new data indicate that the field reversal is actually diagonal rather than vertical across the Galactic disk, opening up the possibility of a current sheet {\it{contained within}} the disk, rather than perpendicular to it. This is contrary to the large-scale analysis of the GMF by Kronberg \& Newton-McGee \citeyearpar{kronberg}, who concluded that no reversals occur across the Galactic plane. However, Fig. 4 of Kronberg \& Newton-McGee does show a slight deviation from the symmetric model in the present region of interest ($52^{\circ}<\ell<72^{\circ}$), which we are able to see in more detail with our much higher density of EG sources per square degree. A current sheet contained within the disk is physically preferable, as there is more material in the disk to support the required current. However, a source or driving mechanism would need to exist in order for such a current to be sustained.

\begin{figure}
  \resizebox{\hsize}{!}{\includegraphics{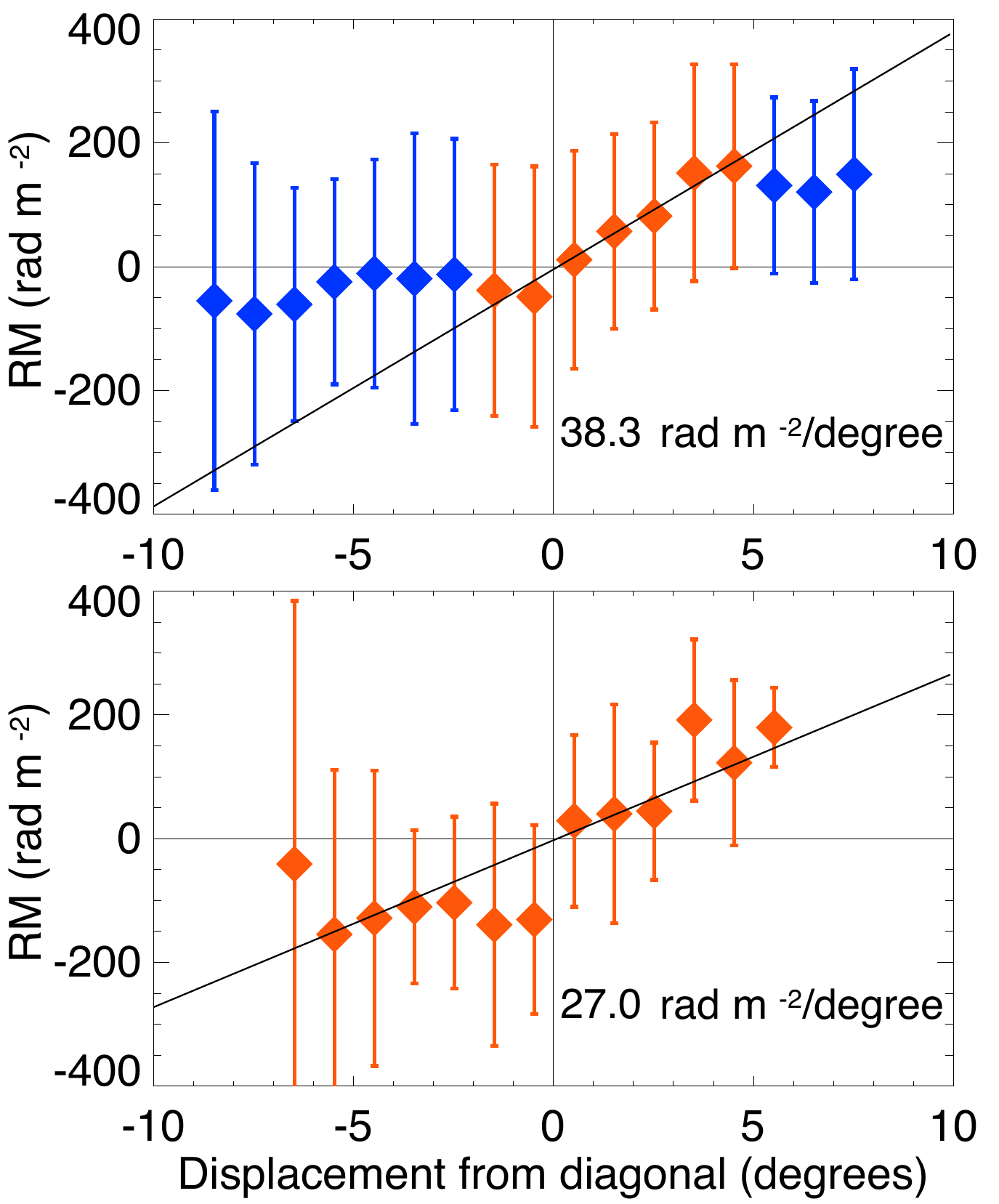}}
  \caption{RM versus displacement from the diagonal boundary in the RM map for XE data (\textit{top panel}) and EG data (\textit{bottom panel}). Positive (negative) displacement corresponds to regions above (below) the boundary. Blue symbols indicate where the RM versus displacement is relatively flat. The slopes quantifying the gradient are calculated using the orange symbols. Note that for the EG data, the slope is more gradual and no clear flat regions are identifiable. The error bars were determined as the standard deviations within the bins.}
  \label{fig3}
\end{figure}

\looseness=-1 To further examine this diagonal reversal, we plot RMs as a function of angular displacement perpendicular to the boundary we identified between the regions of positive and negative RMs, as shown in Fig. \ref{fig3}, with XE and EG RMs averaged into 1$^{\circ}$ bins. For the XE RMs the gradient has a slope of \break 38.3 rad m$^{-2}$ degree$^{-1}$. For an infinitesimally thin current sheet, the gradient would be a step function. Instead, our observations suggest a current ``slab'' of finite thickness, where the slope depends on the thickness and current density of the slab. For the EG RMs the gradient is less steep, with a slope of \break 27.0 rad m$^{-2}$ degree$^{-1}$. This could be attributed to the longer lines of sight for the EG sources. If the current sheet defining the boundary between opposing magnetic field directions is tilted not only in the plane of the sky-projection but is also tilted along the LOS, then RMs resulting from longer lines of sight would have more latitudinal variation, which would cause the gradient to be smeared out instead of sharp. Such an effect could be present in both XE and EG datasets even in the case of a very thin current sheet.

Determining the thickness of the current slab from Amp\`ere's law would require knowledge about the current density in addition to the magnetic field gradient. Assuming a magnetic field strength of 2 $\mu$G on either side of the current slab with opposing directions on the two sides, and a uniform current density, the thickness of the slab multiplied by the current density would be roughly $3\times10^{-4}$ A m$^{-1}$. This corresponds to either a low current density or a thin slab. Observational techniques for determining the thickness of the slab would include comparison of pulsar RMs (with known distances) on and near the identified boundary to EG and XE RMs, along with a more accurate determination of the distance to the PH. Further investigations of the field reversal in this region will involve modelling the current slab with varying thickness, inclination and current density, and fitting such models to the RM datasets.

It is clear that three-dimensional modelling of the GMF is indispensable for describing this diagonal boundary between opposing field directions. A new model that accounts for this type of structure is the hybrid dynamo model of Gressel et al. \citeyearpar{gresseletal}, a three-dimensional global simulation of the disk magnetic field that allows for evolution of the system in contrast to static solutions. This model also includes magneto-rotational instabilities, which can result in vertical undulations of an antisymmetric, azimuthal field component about the Galactic mid-plane. As Gressel et al. point out, this would lead to apparent radial field reversals viewed from near the mid-plane, which would be difficult to detect in external galaxies, particularly if the amplitude of the field undulation is small compared to the scale-height of the Galactic disk, as appears to be the case for the particular model presented in Fig. 10 of Gressel et al. \citeyearpar{gresseletal}. The authors note that it would be worthwhile to investigate whether such a model is consistent with all-sky RM data. Although our present analysis covers only a small segment of the sky, our observations do lend support to the possibility of convective instability-induced, undulating field reversals in the disk.

Another model describing large-scale reversals in the magnetic fields of spiral galaxies is the spiral potential model of Dobbs et al. \citeyearpar{Dobbs}. In these simulations the reversals in the galactic disk are associated with large changes in the velocity field across spiral shocks, as well as changes between inward gas flows along the arms and outward radial flows in inter-arm regions. While this description predicts realistic locations for field reversals, it does not address any latitudinal variation of the reversal location.

\begin{figure}[t]
  \resizebox{\hsize}{!}{\includegraphics{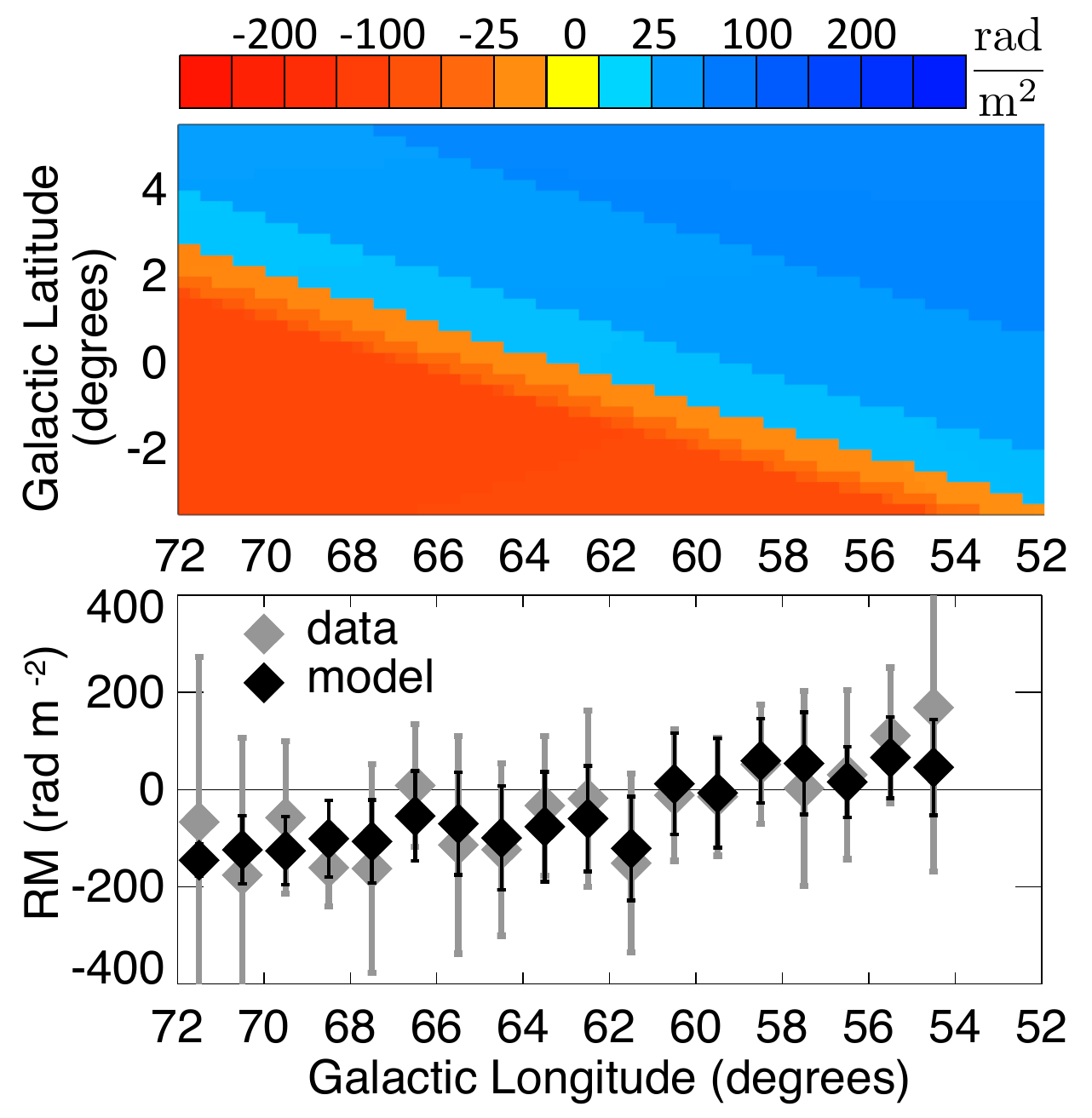}}
  \caption{Results of the Parker spiral-like model for the CGPS low longitude range. \textit{Top panel}: RM map for the XE in the region, showing the diagonal boundary. \textit{Bottom panel}: data (grey) and model (black) RMs averaged into 1$^{\circ}$ longitude bins as a function of Galactic longitude. The error bars in the \textit{bottom panel} were determined as the standard deviations within the bins.}
  \label{fig4}
\end{figure}

An alternative model that could account for the observed diagonal field reversal is the Galactic Parker spiral model of Akasofu \& Hakamada \citeyearpar{akasofuhakamada}, in which a dipolar magnetic field at the Galactic centre is carried outwards by a Galactic wind, analogous to the solar wind. An offset between the axes of the Galaxy and its magnetic dipole would cause a vertical oscillation of azimuthal and radial field components about the mid-plane similar to that described by Gressel et al. \citeyearpar{gresseletal}. The Parker spiral provides a mathematical description that can easily fit our data (see Fig. \ref{fig4}), while also describing other features of the GMF. It explains the absence of observed field reversals in other galaxies, and predicts a decline in the pitch angle of the spiral field pattern  with increasing galactocentric distance, although the latter may also be attributed to flaring of the outer disk \citep{gresseletal,fletcher}.

The Parker model was rejected almost immediately \citep{vallee} but many of the bases for that rejection are no longer valid, given presently available data. The Parker spiral is a convenient mathematical tool, though it does have difficulty generating the flow of material from the Galactic centre, orthogonal to the spiral arms, that is needed to sustain the field in this configuration. Nevertheless, it is a simple model that is able to explain many observed GMF features. Considering similarities between Galactic and solar magnetohydrodynamics is likely to be informative even if the two systems are not exactly alike. Further investigations into a Parker-spiral-based model for the GMF will need to consider how the geometry changes due to sub-virial outflow, the orbital motion of the interstellar medium, and field generation through a Galactic-scale dynamo.

\section{Summary}
We have demonstrated that RMs of diffuse synchrotron emission from within our Galaxy, as observed with the Synthesis Telescope at the DRAO, can provide useful information on large-scale magnetic field structures. Although the XE data, which lack a single-antenna component, miss the largest angular structures, the interferometer is sensitive to high angular frequencies, and hence detects very well sharp changes in the image plane. This allowed us to examine more carefully the well-established magnetic field reversal between our local arm and the Sagittarius Arm. What we found was a diagonal boundary separating positive and negative RMs in the lower longitude region of the CGPS, suggesting the presence of a diagonally oriented current sheet, and highlighting the need for three-dimensional modelling of the GMF, in contrast with two-dimensional models of the disk that can only account for strictly radial reversals. We have noted, as well, the strong resemblance between RM maps derived from EG and XE data, over an extended area, which will require additional investigation, including modelling and comparison to pulsar RMs, as the similarity is not expected due to the difference in length scales being probed. 

Future work will include the addition of broad spatial structure to the RM maps by combining the four band CGPS data with corresponding frequency bands of the Global Magneto-Ionic Medium Survey (Wolleben et al. 2009). New observations complementary to the CGPS at higher latitudes are currently underway to study continuation of the RM gradient above and below the Galactic disk. Understanding the observed diagonal boundary will contribute to more accurate three-dimensional models of the GMF structure.

\begin{acknowledgements}
      We gratefully acknowledge L. Nicolic, T. Foster, B. Jackel, D. Knudsen, B. Gaensler and J. Dickey for enlightening conversations. We thank the anonymous referee whose comments and suggestions have improved this manuscript. The Dominion Radio Astrophysical Observatory is a National Facility operated by the National Research Council Canada. The Canadian Galactic Plane Survey is a Canadian project with international partners, and is supported by the Natural Sciences and Engineering Research Council (NSERC).
\end{acknowledgements}

\bibliographystyle{aa}
\bibliography{ms_aa}

\end{document}